\documentclass[conference]{IEEEtran}
\IEEEoverridecommandlockouts

\usepackage{cite}
\usepackage{amsmath,amssymb,amsfonts}
\usepackage{algorithmic}
\usepackage{graphicx,subfigure}
\usepackage{textcomp}
\usepackage{xcolor}
\usepackage{flushend}
\usepackage{fancyhdr}

\def\BibTeX{{\rm B\kern-.05em{\sc i\kern-.025em b}\kern-.08em
    T\kern-.1667em\lower.7ex\hbox{E}\kern-.125emX}}

\fancypagestyle{firstpage}{
  \setlength{\headheight}{3.5\normalbaselineskip}
  \setlength{\headsep}{1.5\normalbaselineskip}

  \fancyhead[C]{%
    \footnotesize%
    To be presented at the \textit{Picture Coding Symposium} 2025, Aachen, Germany, December 2025 \\
    \vspace{0.8\normalbaselineskip}%
    \scriptsize
    \copyright{} 2025 IEEE. Personal use of this material is permitted. Permission from IEEE must be obtained for all other uses, in any current or future media, including reprinting/republishing this material for advertising or promotional purposes, creating new collective works, for resale or redistribution to servers or lists, or reuse of any copyrighted component of this work in other works.%
  }
  \fancyfoot{}
}

\begin{document}

\title{Bit Allocation Transfer for Perceptual Quality Enhancement of VVC Intra Coding 
\thanks{This work was supported in part by the Natural Sciences and Engineering Research Council (NSERC) of Canada.}
}

\author{\IEEEauthorblockN{Runyu Yang}
\IEEEauthorblockA{\textit{School of Engineering Science} \\
\textit{Simon Fraser University}\\
Burnaby, BC, Canada \\
runyuy@sfu.ca}
\and
\IEEEauthorblockN{{Ivan V. Baji\'c}}
\IEEEauthorblockA{\textit{School of Engineering Science} \\
	\textit{Simon Fraser University}\\
	Burnaby, BC, Canada \\
ibajic@ensc.sfu.ca}
}

\maketitle

\begin{abstract}

Mainstream image and video coding standards -- including state-of-the-art codecs like H.266/VVC, AVS3, and AV1 -- adopt a block-based hybrid coding framework. While this framework facilitates straightforward optimization for Peak Signal-to-Noise Ratio (PSNR), it struggles to effectively optimize perceptually-aligned metrics such as Multi-Scale Structural Similarity (MS-SSIM). To address this challenge, this paper proposes a low-complexity method to enhance perceptual quality in VVC intra coding by transferring bit allocation knowledge from end-to-end image compression. We introduce a lightweight model trained with perceptual losses to generate a quantization step map. This map implicitly captures block-level perceptual importance, enabling efficient derivation of a QP map for VVC. Experiments on Kodak and CLIC datasets demonstrate significant advantages, both in execution time and perceptual metric performance, with more than 11\% BD-rate reduction in terms of MS-SSIM. Our scheme provides an efficient, practical pathway for perceptual enhancement of traditional codecs.

\end{abstract}


\thispagestyle{firstpage}


\section{Introduction}

Contemporary image and video compression standards like H.266/Versatile Video Coding (VVC) \cite{bross2021developments}, AVS3 \cite{zhang2019recent}, and AV1 \cite{chen2018overview}—universally adhere to a block-based hybrid coding framework. A fundamental challenge in designing an encoder for any such standard is achieving the optimal compromise between compression efficiency (bitrate) and reconstruction quality. Formally, this is the Rate-Distortion Optimization (RDO) problem:
\begin{equation}\label{equ:RDO}
	\min D \quad \text{subject to} \quad R \leq R_c,
\end{equation}
where $D$ represents the distortion (inversely related to quality), $R$ is the actual bitrate consumed, and $R_c$ is a target bitrate constraint. Block-based coding inherently facilitates granular RDO decisions at the block level. The encoder makes critical choices for each block: selecting the optimal prediction mode, determining the best partition size, and choosing a suitable Quantization Parameter (QP) value. This localized decision-making flexibility is a cornerstone of the high compression efficiency achieved by modern standards like VVC.

\begin{figure}
	\begin{minipage}[b]{1.0\linewidth}
		\centering
		\centerline{\includegraphics[width=\linewidth]{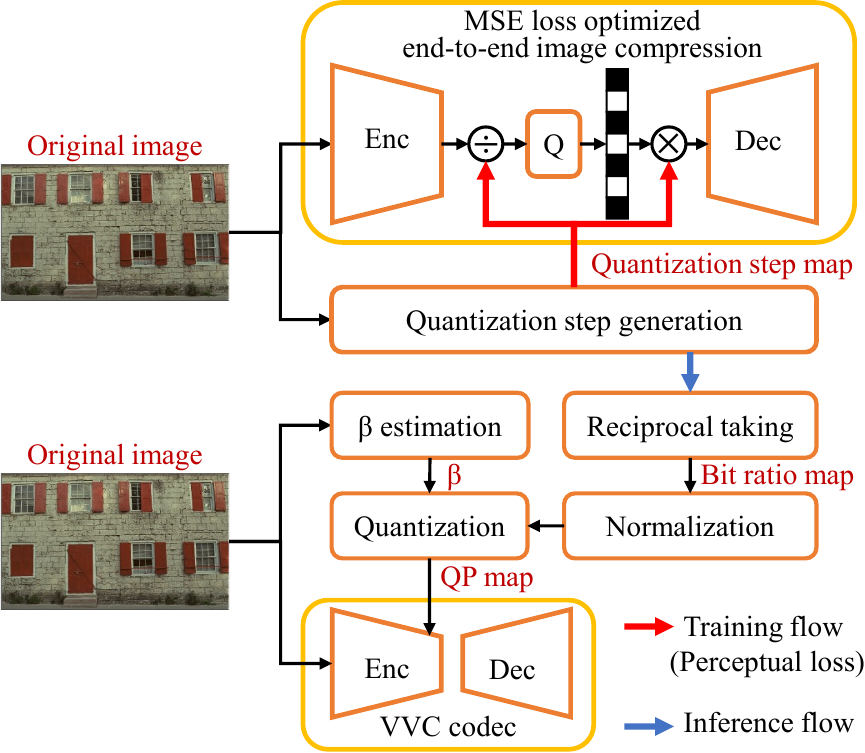}}
	\end{minipage}
	\caption{The proposed VVC intra coding scheme with perceptual quality enhancement via transfering from end-to-end image compression bit allocation.}
	\label{fig:overall}
\end{figure}

Crucially, the definition and calculation of distortion ($D$) significantly impact optimization outcomes. The Mean Squared Error (MSE), and its derived Peak Signal-to-Noise Ratio (PSNR), have been ubiquitous metrics due to their mathematical simplicity and tractability. However, decades of psychovisual research have established a critical limitation: MSE/PSNR correlates poorly with human perceptual quality \cite{girod1993s,wang2009mean}. Consequently, significant research efforts have yielded perceptually-driven metrics designed to better align with human vision. These include the Structural Similarity Index (SSIM) \cite{wang2004image}, its multi-scale extension MS-SSIM \cite{wang2003multiscale}, and modern data-driven metrics like the Learned Perceptual Image Patch Similarity (LPIPS) \cite{zhang2018unreasonable}. While demonstrably superior to MSE in correlating with subjective quality, these metrics lack the additive separability and block independence inherent in MSE. Optimizing the RDO trade-off block-by-block using complex, non-additive perceptual metrics like MS-SSIM or LPIPS is computationally complex and often infeasible within the standard framework, limiting their practical adoption despite their alignment with perceptual notions of visual quality.

\begin{figure}
	\begin{minipage}[b]{1.0\linewidth}
		\centering
		\centerline{\includegraphics[width=\linewidth]{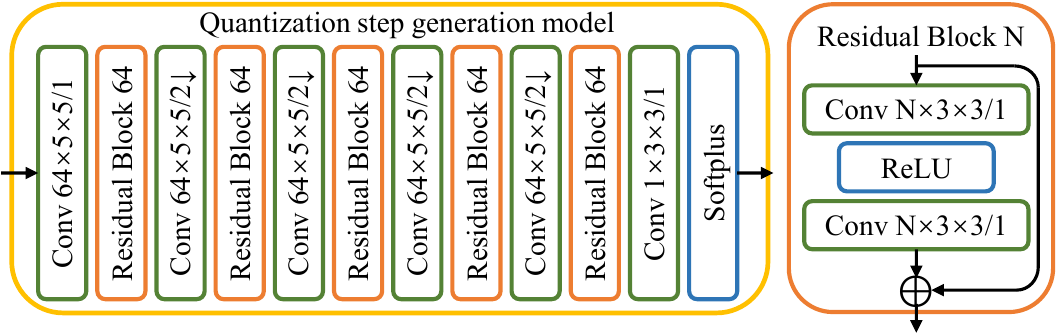}}
	\end{minipage}
	\caption{Left: Quantization step generation model; Right: Residual Block.}
	\label{fig:ada_model}
\end{figure}

Recently, end-to-end image compression has emerged as a powerful alternative to conventional compression~\cite{balle2018variational,wu2021learned,ma2020end,he2022elic,liu2023learned}. This paradigm utilizes deep neural networks to implement the entire encoder-decoder pipeline. Its approach to RDO is fundamentally different: optimization occurs globally during offline training over a large dataset of natural images. A critical advantage is metric flexibility. As long as the distortion metric is differentiable with respect to the network parameters, it can be seamlessly incorporated into the gradient-based training process. This makes employing perceptually-aligned metrics like MS-SSIM or LPIPS much easier than with conventional codecs. For a given input image, networks optimized for different metrics exhibit significantly different patterns of bit allocation across spatial regions (blocks). This implicitly learned block-wise bit allocation reflects the understanding of the relative perceptual importance of different image areas under the specific target metric.

This leads to a core research question: Can the implicit knowledge of perceptual block importance, learned by end-to-end models, be extracted and transferred to guide RDO decisions within traditional block-based encoders? Some works~\cite{yang2021knowledge,yang2024perceptual} have demonstrated the feasibility of this concept. They extracted bit allocation patterns from both an MS-SSIM-optimized and an MSE-optimized end-to-end image compression model, integrating this knowledge into traditional block-based encoders by bit adaptation. While this approach showed promising initial gains in MS-SSIM performance, it requires running the encoder network twice, resulting in high computational complexity. This paper expands upon that foundational concept, presenting a low-complexity scheme that uses a lightweight model to generate quantization steps for the compression model, and then transfers this information to a VVC encoder. This approach significantly reduces complexity while maintaining performance.

The rest of this paper is organized as follows. Section \ref{sec:Method} describes the proposed perceptual quality enhancement scheme. Section \ref{sec:Experiments} provides the detailed experimental results. Section \ref{sec:Discussion} discusses the future work of the scheme. Finally, Section \ref{sec:Conclusion} concludes this paper.

\section{Proposed method}
\label{sec:Method}

The overall scheme of our method is introduced in Section \ref{sec:Overall}, followed by detailed descriptions of various components in subsequent sections.

\subsection{Overall scheme}
\label{sec:Overall}

Fig.~\ref{fig:overall} illustrates the overall bit allocation transfer scheme. Our objective is to generate a QP map for block-based image compression like VVC to improve perceptual quality. First, we train the quantization step generation model with a perceptual loss. This generates a quantization step map. The reciprocal of this map approximates the bit ratio. We take this reciprocal, normalize it for rate alignment, and derive a bit ratio map. Finally, we quantize this ratio map into the target QP map via the R-$\lambda$ model \cite{li2014lambda}. Finally, VVC uses this QP map to adaptively allocate bits for perceptual enhancement.

\begin{table}
	\centering
	\caption{Average BD-rate results (\%) of different metrics than MSE optimized NIC on the Kodak}
	\begin{tabular}{c|c|c|c|c}
		\hline
		Method & PSNR       & SSIM    & MS-SSIM      & LPIPS  \\ \hline
		NIC (SSIM)    & 2.44    & $-$6.89    & $-$6.44     & $-$9.87    \\ \hline
		NIC (MS-SSIM) & 4.56   & $-$5.62   & $-$7.21   & $-$8.94   \\ \hline
		NIC (LPIPS)   & 0.31  & $-$2.89   & 4.58  & $-$9.73  \\ \hline
	\end{tabular}
	\label{table:performance}
\end{table}

\subsection{End-to-end image compression}

Our scheme is model-agnostic, requiring only that the end-to-end compression architecture permits block-level bit allocation. We selected NIC\footnote{https://github.com/fvc-sg/NIC} for implementation which follows mainstream hyper-prior design \cite{balle2018variational}. It minimizes the joint loss:
\begin{equation}\label{equ:loss function}
	L = \lambda D + R_{main} + R_{hyper}.
\end{equation}
Here, $\lambda$ controls the rate-distortion trade-off, $D$ is distortion defined by MSE or perceptual metrics, while $R_{main}$ and $R_{hyper}$ correspond to bitrates of quantized latent features and hyperpriors (the latter providing side information for entropy coding). The latent features have a spatial resolution downsampled by a factor of 16 relative to the original image.

\subsection{Quantization step generation}

The bitrate required for encoding quantized latent features is determined by their magnitude – larger values probably mean higher bitrates, analogous to quantized coefficients in VVC. To adaptively control this bit allocation, we employ a quantization step generation model that produces a quantization step map. This map inversely regulates the magnitude of the quantized latent features: step values larger than one reduce feature magnitudes and bitrates, while steps smaller than one increase them. As shown in Fig.~\ref{fig:ada_model}, the model architecture comprises stacked residual blocks and stride-2 convolutional layers. The resolution of its output is the same with the latent feature. We use softplus as the final activation function to ensure positive outputs suitable for this application. The softplus is defined as
\begin{equation}\label{equ:softplus}
\text{softplus}(x) = \ln(1+e^x).
\end{equation}

During training, we fix the image compression model and train only the quantization step generation model. Table~\ref{table:performance} demonstrates improved performance metrics under this configuration (the bitrate does not contain the quantization step map). These gains confirm that the generated quantization step map effectively distills perceptual block importance knowledge from the compression model.

	\begin{figure}
		\centering
		\subfigure[Kodak]{
		\begin{minipage}[b]{0.48\linewidth}
	\includegraphics[width=1\linewidth]{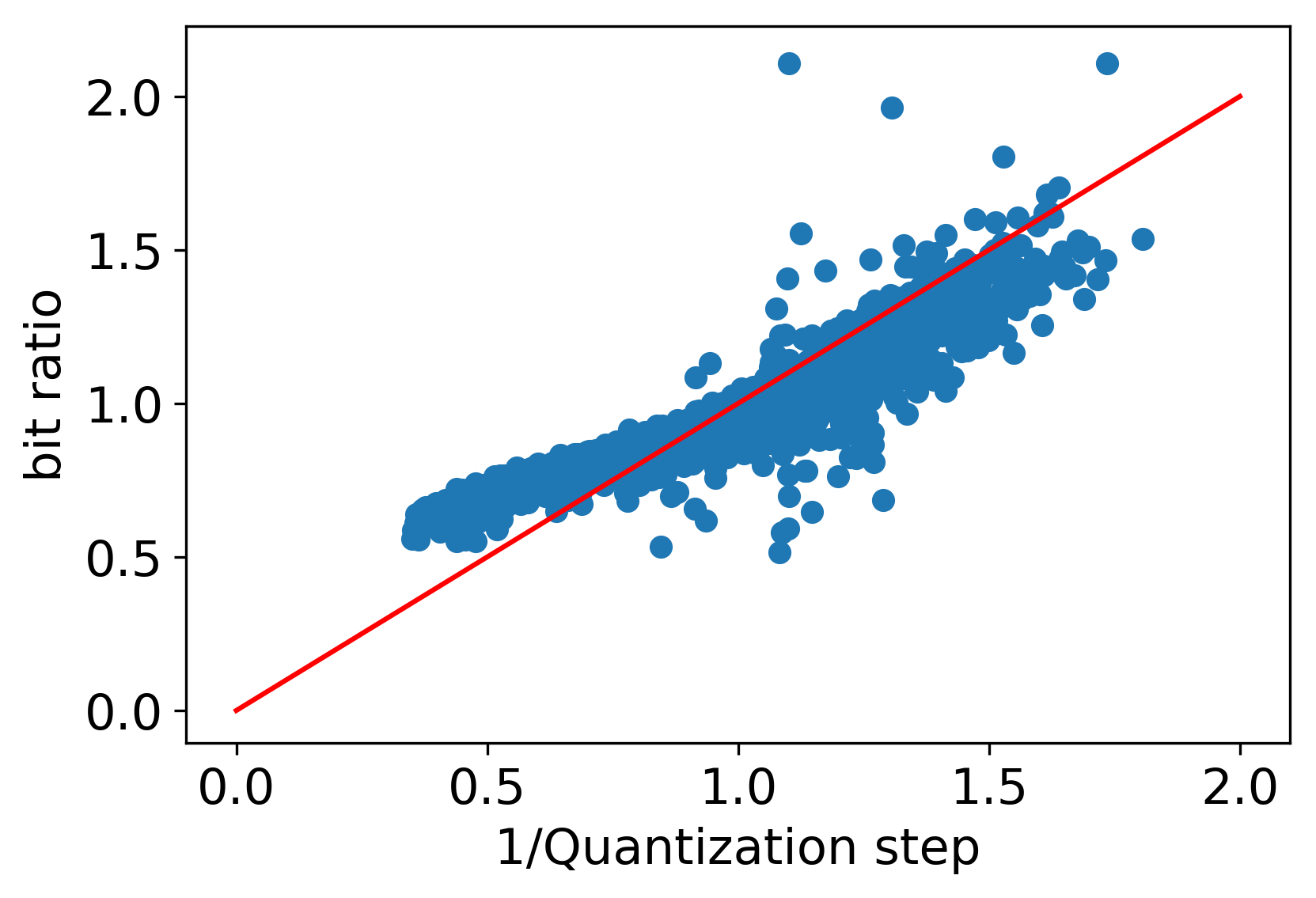}
    \end{minipage}}
\subfigure[CLIC]{
		\begin{minipage}[b]{0.48\linewidth}
	\includegraphics[width=1\linewidth]{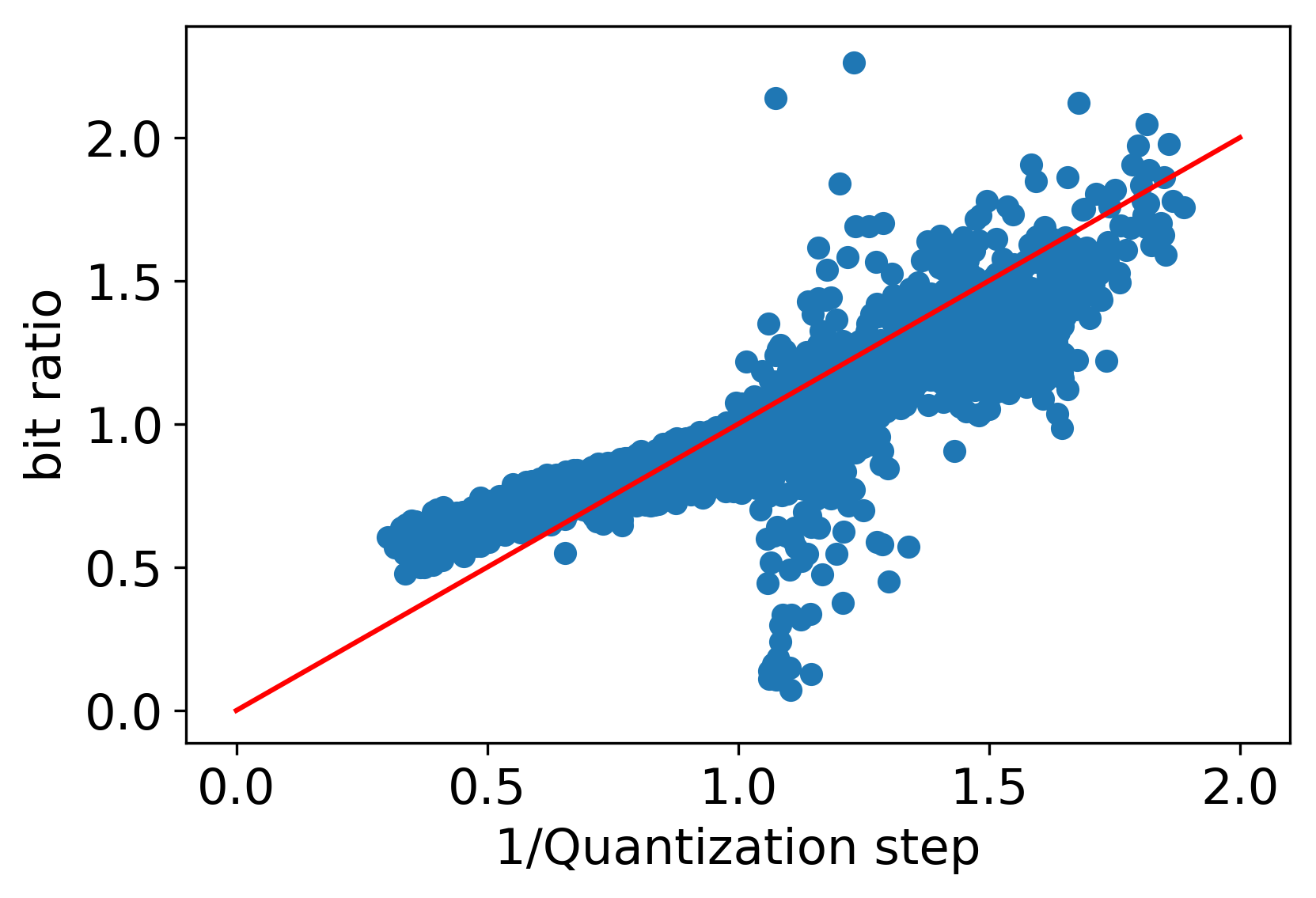}
\end{minipage}}
		\caption{The relationship between bit ratio and the reciprocal of the quantization step on Kodak and CLIC. The red line indicates a linear relationship with a slope of 1.}
		\label{fig:QS}
	\end{figure}

\subsection{QP adaptation}

	To allocate bits of each block according to the quantization step map, the quantization step map should be quantified to the QP map. We follow the method used in \cite{yang2024perceptual} which needs the block-based bit adaptation ratio. We found that bit ratio scales linearly with the reciprocal of the quantization step in Fig.~\ref{fig:QS} (granularity of 64$\times$64 block, take the average step of 4$\times$4), as demonstrated by the red line with a slope equal to 1. Therefore, the bit ratio can be approximated as the reciprocal of the quantization step. The bit ratio is
	\begin{equation}\label{equ:ratio}
		r_k \approx \frac{1}{QS_k},
	\end{equation}
	where $k$ is the block index, $r_k$ is the block-based bit ratio for guiding the bit allocation, $QS_k$ is the block-based quantization step, which takes the average value within the block position of the quantization step map. 
	
	To determine the rate $R$ by the Lagrange multiplier $\lambda$, we adopt the R-$\lambda$ model \cite{li2014lambda}. The basic R-$\lambda$ model reads
		\begin{equation}\label{equ:R-lambda}
		\lambda = \alpha \cdot R^\beta,
	\end{equation}
	where $\alpha$ and $\beta$ are hyperparameters and vary with block content. $\lambda$ and $R$ are both frame-wise values. Because we adapt QP based on block while $\alpha$ and $\beta$ vary between blocks, so we can get
	\begin{equation}\label{equ:R-lambda2}
		\lambda = \alpha_k \cdot R_k^{\beta_k},
	\end{equation}
	where $k$ is the block index. To achieve bit allocation based on the block-based bit ratio $r_k$, the block-wise $\lambda_k$ is
	\begin{align}\label{equ:R-lambda3}
		\lambda_k &= \alpha_k \cdot \left({r_k}{R_k}\right)^{\beta_k}\\
		&= {r_k}^{\beta_k} \cdot \alpha_k \cdot R_k^{\beta_k}\\
		&= {r_k}^{\beta_k} \cdot \lambda.
	\end{align}

	According to the relationship between QP and $\lambda$ in VVC, the QP value for the $k$-th block is,
	\begin{align}\label{equ:QP2}
		\text{QP}_k &= \text{QP} + N\log_2 \left(\frac{\lambda_k}{\lambda}\right) \\
		&= \text{QP} + 3\log_2 ({r_k}^{\beta_k})\\
		&\approx \text{QP} - 3\log_2 ({QS_k}^{\beta_k}),
	\end{align}
where $N$ is set to 3 according to VVC settings. The corresponding $\lambda$ value for the $k$-th block, $\lambda_k$, also needs to be adapted as:
	\begin{equation}\label{equ:lambda}
		\lambda_k = \lambda \cdot 2^{\left({\text{QP}_k-\text{QP}}\right)/3}.
	\end{equation}

\section{Experiments}
\label{sec:Experiments}

\subsection{Settings}

Our training process and model for $\beta_k$ estimation are the same as \cite{yang2024perceptual,li2017convolutional}. Our quantization step generation training data originates from the LIU4K dataset \cite{liu2020comprehensive}, comprising 607,714 randomly cropped 256$\times$256 patches extracted from 1,600 original images and their 2$\times$/4$\times$ bicubic-downsampled versions.

We benchmark against VTM-23.0, Distillation \cite{yang2024perceptual}, and PerceptQPA-enabled VTM-23.0 \cite{bosse2017Perceptually,helmrich2018Improved} – the latter employing high-pass filtering for QP adaptation. For compatibility with VTM-23.0, all images were converted to YUV420 format. To ensure fair comparison with PerceptQPA (which additionally allocates bits to chroma components), we maintained its 64$\times$64 QP adaptation block size while replacing only its block-based QP adaptation module. A maximum QP offset of 4 was enforced to mitigate blocking artifacts. Rate alignment with VTM-23.0 was achieved by training models at QP $\in \{37, 32, 27, 22\}$.

Our joint rate-distortion loss function is defined as:
\begin{equation}\label{equ:loss function2}
	L = \lambda (\alpha D_{perc}) + R_{main} + R_{hyper},
\end{equation}
where $\lambda$ and $\alpha$ balance the trade-off between rate and distortion. We train three perceptual variants: 1-SSIM, 1-MS-SSIM, and LPIPS. The LPIPS metric sums feature-space distances across five convolutional layers of an AlexNet-based network originally adapted from image classification. The hyperparameter $\alpha$ is set to 0.02, 0.08, and 0.04 for each loss type, respectively.

The end-to-end image compression model used pre-trained MSE-optimized weights and trained the quantization step generation model for five epochs with Adam, using initial learning rate of 1e-4. Rate alignment used $\lambda \in \{1,4,8,16\}$ corresponding to $\text{QP} \in \{37, 32, 27, 22\}$.

Evaluation metrics include RGB PSNR, SSIM, MS-SSIM, and LPIPS. For BD-rate calculation, LPIPS values were converted to dB scale:
\begin{equation}\label{equ:LPIPS}
	\text{LPIPS (dB)} = -10\log_{10}(\text{LPIPS}).
\end{equation}

Our experiments employ two datasets: the Kodak image set\footnote{http://r0k.us/graphics/kodak} (24 images, about 0.35MP) and CLIC 2022 validation/test images\footnote{http://www.compression.cc} (more than 2MP). Since VTM-23.0 requires image dimensions divisible by 4, we select compliant CLIC images. Additionally, to control for PerceptQPA's chroma component bit allocation, we test a zero QP map configuration where quantization parameters remain static.

	\begin{table}
		\centering
		\caption{Average BD-rate results (\%) of our method, Distillation, and PerceptQPA, compared to VTM-23.0 on the \textbf{Kodak}}
		\begin{tabular}{c|cccc}
			\hline
			Method & PSNR & SSIM & MS-SSIM & LPIPS \\ \hline
			Zero QP map & $-$2.75 & $-$1.04 & $-$3.46 & $-$2.42 \\ 
			PerceptQPA & 2.85 & $-$4.26 & $-$11.86 & $-$11.96 \\ \hline
			Distillation opt. SSIM & 1.80 & $-$6.14 & $-$10.11 & $-$11.98 \\
			Ours opt. SSIM & $-$0.20 & $-$5.71 & $-$10.05 & $-$8.80 \\\hline
			Distillation opt. MS-SSIM & 2.52 & $-$5.83 & $-$12.74 & $-$13.30 \\
			Ours opt. MS-SSIM & 0.98 & $-$6.19 & $-$11.88 & $-$10.96 \\\hline
			Distillation opt. LPIPS & 0.65 & $-$3.66 & $-$10.28 & $-$11.00 \\
			Ours opt. LPIPS & $-$0.10 & $-$2.90 & $-$8.55 & $-$8.33 \\
			\hline
		\end{tabular}
		\label{table:result1}
	\end{table}

\subsection{Results of different optimization objectives}

Table~\ref{table:result1} presents the BD-rate results for the Kodak image set, while Table~\ref{table:result2} shows those for the CLIC dataset. All evaluated methods demonstrate improvements in perceptual metrics. Our method, optimizing SSIM and MS-SSIM, outperforms PerceptQPA on the corresponding perceptual metrics. Although our approach exhibits a slight disadvantage in these perceptual metrics compared to Distillation \cite{yang2024perceptual}, it achieves superior PSNR retention. Notably, on the CLIC dataset, our method surpasses both PerceptQPA and \cite{yang2024perceptual} in both perceptual metrics (SSIM and MS-SSIM) and PSNR. This enhanced performance stems from our method's tendency towards conservative QP adaptation, which benefits high-resolution images.

However, our LPIPS-optimized variant showed no significant advantages, aligning with the findings in \cite{yang2024perceptual}. This occurs because optimizing deep perceptual metrics amplifies the performance gap between block-based image compression (like VVC) and end-to-end image compression, potentially compromising final results.

To explore more aggressive QP adaptation within our scheme, we increased the slope size to 1.2 shown in Table~\ref{table:result3}. This adjustment improved perceptual metrics at the cost of reduced PSNR performance. Under this configuration, SSIM and MS-SSIM performance on Kodak became comparable to Distillation \cite{yang2024perceptual}.

Furthermore, we assessed the effectiveness of approximating the bit ratio using the reciprocal of the quantization step size by comparing it against the true bit ratio method used in Distillation \cite{yang2024perceptual} (Table~\ref{table:result4}). Results indicate that the reciprocal approximation enables more aggressive QP adaptation, leading to greater gains in perceptual metric performance.

\begin{figure*}
\begin{minipage}[b]{1\linewidth}
\centering
\centerline{\includegraphics[width=1\linewidth]{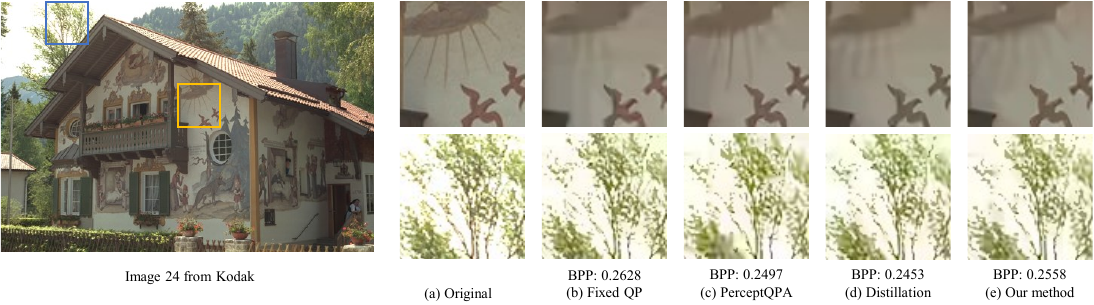}}
\end{minipage}
\caption{Visual evaluation of our method optimizing MS-SSIM under QP 37.}
		\label{fig:visual}
\end{figure*}

Fig.~\ref{fig:visual} shows the visual evaluation of our method optimizing MS-SSIM under QP 37. Compared to VTM-23.0 (fixed QP), at regions with structure, our method improves the perceptual quality; at highly textured regions, our method produces similar perceptual quality at a lower rate. It is similar to PerceptQPA and Distillation.

\subsection{Computational complexity}

Regarding time complexity, we measured the QP map generation time on a single PC running Windows 11 64-bit, equipped with an AMD Ryzen 7 5800H CPU, 16 GB of RAM, and an NVIDIA GeForce GTX 1080 Ti GPU, using PyTorch 1.5.1. By employing a lightweight model for quantization step generation instead of running the encoder network twice, our QP map generation is significantly faster than Distillation \cite{yang2024perceptual}, costing only around 20 ms and 700 ms on GPU and CPU, respectively, for a Kodak image. This represents less than one-tenth of the time using Distillation\cite{yang2024perceptual}. Although this time constitutes only a small fraction of the overall VVC encoding process, it remains non-negligible in practical applications.

	\begin{table}
		\centering
		\caption{Average BD-rate results (\%) of our method, Distillation, and PerceptQPA, compared to VTM-23.0 on the \textbf{CLIC}}
		\begin{tabular}{c|cccc}
			\hline
			Method & PSNR & SSIM & MS-SSIM & LPIPS \\ \hline
			Zero QP map & $-$2.63 & $-$0.21 & $-$2.46 & $-$1.21 \\ 
			PerceptQPA & 3.20 & $-$2.42 & $-$9.91 & $-$11.51 \\ \hline
			Distillation opt. SSIM & 7.39 & $-$4.80 & $-$7.30 & $-$11.15 \\
			Ours opt. SSIM & 0.57 & $-$5.45 & $-$9.53 & $-$8.33 \\\hline
			Distillation opt. MS-SSIM & 7.55 & $-$3.61 & $-$10.24 & $-$11.97 \\
			Ours opt. MS-SSIM & 2.46 & $-$5.91 & $-$11.26 & $-$10.88 \\\hline
			Distillation opt. LPIPS & 4.44 & $-$2.96 & $-$9.68 & $-$10.65\\
			Ours opt. LPIPS & 1.10 & $-$1.65 & $-$7.12 & $-$6.11 \\
			\hline
		\end{tabular}
		\label{table:result2}
	\end{table}

\section{Discussion}
\label{sec:Discussion}

While the proposed scheme demonstrates efficient perceptual enhancement for VVC intra coding, several promising directions warrant future exploration. First, adopting end-to-end image compression architectures that better align with traditional coding frameworks (e.g., block-based structures \cite{wu2021learned}) could bridge the performance gap for metrics like LPIPS. Such architectural synergy may improve knowledge transferability and enhance optimization for deep perceptual metrics. Second, extending this approach to video coding is critical. Temporal consistency could be addressed by transferring bit allocation knowledge from end-to-end video compression models, potentially leveraging motion-guided perceptual importance to refine QP adaptation across frames. Third, optimizing bit allocation for machine vision tasks represents an emerging frontier. By training the quantization step generaiton model with task-specific losses (e.g., object detection accuracy), the scheme could dynamically allocate bits to regions critical for downstream algorithms. Future work will explore these avenues to broaden the applicability and impact of this transfer-based bit allocation scheme.

\section{Conclusion}
\label{sec:Conclusion}

This paper presented a low-complexity scheme for enhancing the perceptual quality of VVC intra coding by transferring bit allocation from end-to-end image compression. We first introduced a lightweight quantization step generation model trained with perceptual losses to distill block-level importance information. This quantization step map is then efficiently converted into a bit ratio map using a reciprocal approximation. At last, the QP map is generated by the bit ratio map for the perceptual quality enhancement of VVC. Compared to prior methods requiring dual encoder runs, our approach significantly reduces QP map generation time, making it practical. Experiments demonstrate that our method, particularly when optimizing SSIM and MS-SSIM, outperforms PerceptQPA and achieves comparable or superior perceptual quality to Distillation \cite{yang2024perceptual} while offering better PSNR retention. Our scheme provides an effective low-complexity pathway for perceptual quality enhancement in traditional codecs.

	\begin{table}
		\centering
		\caption{Average BD-rate (\%) of our method with the variant of slope equal to 1.2, compared to VTM-23.0 on the Kodak dataset}
		\begin{tabular}{c|cccc}
			\hline
			Method & PSNR & SSIM & MS-SSIM & LPIPS \\ \hline
			opt. SSIM & $-$0.20 & $-$5.71 & $-$10.05 & $-$8.80 \\
			opt. SSIM (slope = 1.2) & 0.82 & $-$6.16 & $-$10.86 & $-$9.75 \\	\hline
			opt. opt. MS-SSIM & 0.98 & $-$6.19 & $-$11.88 & $-$10.96 \\
			opt. MS-SSIM (slope = 1.2)& 2.24 & $-$6.28 & $-$12.47 & $-$11.58 \\	\hline
			opt. LPIPS & $-$0.10 & $-$2.90 & $-$8.55 & $-$8.33 \\
			opt. LPIPS (slope = 1.2) & 1.58 & $-$2.02 & $-$8.28 & $-$8.79 \\ \hline
		\end{tabular}
		\label{table:result3}
	\end{table}

	\begin{table}[t]
		\centering
		\caption{Average BD-rate (\%) of our method with the variant adopting real bit ratio, compared to VTM-23.0 on the Kodak dataset}
		\begin{tabular}{c|cccc}
			\hline
			Method & PSNR & SSIM & MS-SSIM & LPIPS \\ \hline
			opt. SSIM & $-$0.20 & $-$5.71 & $-$10.05 & $-$8.80 \\
			opt. SSIM (real ratio) & $-$1.87 & $-$4.07 & $-$7.57 & $-$6.65\\	\hline
			opt. opt. MS-SSIM & 0.98 & $-$6.19 & $-$11.88 & $-$10.96 \\
			opt. MS-SSIM (real ratio)& $-$0.99 & $-$4.97 & $-$9.80 & $-$8.92 \\	\hline
			opt. LPIPS & $-$0.10 & $-$2.90 & $-$8.55 & $-$8.33 \\
			opt. LPIPS (real ratio) & $-$1.16 & 0.05 & $-$4.86 & $-$4.79 \\ \hline
		\end{tabular}
		\label{table:result4}
	\end{table}

\bibliographystyle{IEEEtran}
\bibliography{refs}

\end{document}